\documentstyle[epsf,12pt]{article}
\begin{document}
\begin{center}
\Large{\bf Beam Energy Scan a Case for the Chiral Magnetic Effect in 
Au-Au Collisions.}\\
\large{R.S. Longacre$^a$\\
$^a$Brookhaven National Laboratory, Upton, NY 11973, USA}
\end{center}
 
\begin{abstract}
The Chiral Magnetic Effect (CME) is predicted for Au-Au collisions at RHIC. 
However many backgrounds can give signals that make the measurement hard to 
interpret. The STAR experiment has made measurements at different collisions
energy ranging from $\sqrt{s_{NN}}$=7.7 GeV  to 62.4 GeV. In the analysis that
is presented we show that the CME turns on with energy and is not present in
central collisions where the induced magnetic is small.
\end{abstract}
 
\section{Introduction} 

Topological configurations should occur in the hot Quantum Chromodynamic (QCD) 
vacuum of the Quark-Gluon Plasma (QGP) which can be created in heavy ion 
collisions. These topological configurations form domains of local strong 
parity violation (P-odd domains) in the hot QCD matter through the so-called 
sphaleron transitions. The domains might be detected using the Chiral Magnetic
Effect (CME)\cite{warringa} where the strong external magnetic 
field(electrodynamic) at the early stage of a collision(non-central), through 
the sphaleron transitions which induces a charge separation along the direction
of the magnetic field perpendicular to the reaction plane. Such an out
of plane charge separation, however, varies its orientation from event to 
event, either parallel or anti-parallel to the magnetic field (sphaleron or 
antisphaleron). Also the magnetic field can be up or down with respect to the 
reaction plane depending if the ions pass in a clockwise or anti-clockwise 
manner. Any P-odd observable will vanish and only the variance of such 
observable may be detected.

The STAR collaboration\cite{STARCME} has published a measurement of charge 
particle azimuthal correlations consistent with CME expectations. In 
Ref.\cite{voloshin} and used by STAR the CME can be indirectly approached
through a two-particle azimuthal correlation given by
\begin{equation}
\gamma = \langle cos(\phi_1 + \phi_2 - 2\Psi_{RP}) \rangle = \langle cos(\phi_1-\Psi_{RP}) cos(\phi_2-\Psi_{RP}) \rangle - \langle sin(\phi_1-\Psi_{RP}) sin(\phi_2-\Psi_{RP}) \rangle ,
\end{equation}
where $\Psi_{RP}$, $\phi_1$, $\phi_2$ denote the azimuthal angles of the
reaction plane, produced particle 1, and produced particle 2. This two
particle azimuthal correlation measures the difference between the in plane
and out of plane projected azimuthal correlation. If we would rotate all events
such that $\Psi_{RP}$ = 0.0,
then $\gamma$ would become
\begin{equation}
\gamma = \langle cos(\phi_1 + \phi_2) \rangle = \langle cos(\phi_1) cos(\phi_2) \rangle - \langle sin(\phi_1) sin(\phi_2) \rangle  .
\end{equation}
The CME predicts that $\gamma$ $>$ 0 for opposite sign-pairs and $\gamma$ $<$ 0
for same sign-pairs. There are other two particle azimuthal correlation effects
that can depend on the reaction plane driven by elliptic flow even though the
underlying correlation may be independent of the reaction plane. These 
backgrounds are summarized in Ref.\cite{koch}. It was pointed in 
Ref.\cite{koch} that the $\phi$ difference correlation ($\delta$) which is 
independent of the reaction plane gives a constraint on the CME and 
backgrounds.
\begin{equation}
\delta = \langle cos(\phi_1 - \phi_2) \rangle = \langle cos(\phi_1) cos(\phi_2) \rangle + \langle sin(\phi_1) sin(\phi_2) \rangle  .
\end{equation}

In Ref.\cite{koch} Transverse Momentum Conservation (TMC) is derived and
demonstrated that if there is no other correlation in the data except
elliptic flow TMC will give a negative
$\langle cos(\phi_1 + \phi_2 - 2\Psi_{RP}) \rangle$ ($\gamma$) which is
$v_2$ smaller than $\langle cos(\phi_1 - \phi_2) \rangle$ ($\delta$).
$\delta$ is a negative number given by TMC and scales as 1/N (N is the number
of particles). Also in Ref.\cite{koch} Local Charge Conservation (LCC) is 
another important background and the details of LCC is found in 
Ref.\cite{pratt}. The authors of Ref.\cite{pratt} point out that for
same sign pairs LCC should give a small negative sign, while for opposite sign
pairs LCC should give a large positive correlation. Using the same coupling
effect to the reaction plane as TMC, one should expect
$\langle cos(\phi_1 + \phi_2 - 2\Psi_{RP})_{+-} \rangle$ ($\gamma$) equal $v_2$
times $\langle cos(\phi_1 - \phi_2)_{+-} \rangle$ ($\delta$) for the LCC.

Ref.\cite{warringa} has pointed out that P-odd domains on the surface of the
fireball omit same charge sign particles in the direction of the magnetic 
field. The particles that escape the surface would be of the same sign while 
the charge particles moving in the opposite direction would be of opposite 
sign. These particles would run into the fireball and be thermalized and loss
their direction (quenched). This effect will be taken into account in the 
latter part of this report.

The paper is organized in the following manner:

Sec. 1 is the introduction to correlations. Sec. 2 presents the 
STAR\cite{STARBES} correlation data which will be used in the analysis. Sec. 3 
separates the correlations up into reaction plane dependent and independent
parts with a further separation into charge dependent and charge independent
amplitudes. Sec. 4 introduce ratios of charge dependent to total amplitudes.
Sec. 5 brings the idea of quenching into the reaction plane dependent analysis.
Sec. 6 presents the summary and discussion.

\section{$\gamma$ and $\delta$ from the Beam Energy Scan at RHIC} 

In this analysis we use STAR\cite{STARBES} data coming from charged particles 
produced in Au-Au collisions at RHIC. 8M $\sqrt{s_{NN}}$= 62.4 GeV (2005),
100M at 39.0 GeV (2010), 46M at 27.0 GeV (2011), 20M at 19.6 GeV (2011), 10M at
11.5 GeV (2010) and 4M at 7.7 GeV (2010) were used.  $\gamma$ 
($\langle cos(\phi_1 + \phi_2 - 2\Psi_{RP}) \rangle$) and $\delta$
($\langle cos(\phi_1 - \phi_2) \rangle$) were extracted for like and unlike 
sign charged pairs as a function of centrality. Figure 1 - 4 show scatter plots
of the correlation data. We have plotted centrality vs log($\sqrt{s_{NN}}$). 
In the plots the label E is equal to $\sqrt{s_{NN}}$ and the most central 
collisions are at 10\% while most peripheral are at 70\%. The vertical scale 
is measured in units of $10^{-4}$. Figure 5 show the average $v_2$ plotted in 
the same type of scatter plot. This $v_2$ is needed to extract the reaction 
plane dependent and independent correlations which we define in Sec. 3.

\begin{figure}
\begin{center}
\mbox{
   \epsfysize 8.0in
   \epsfbox{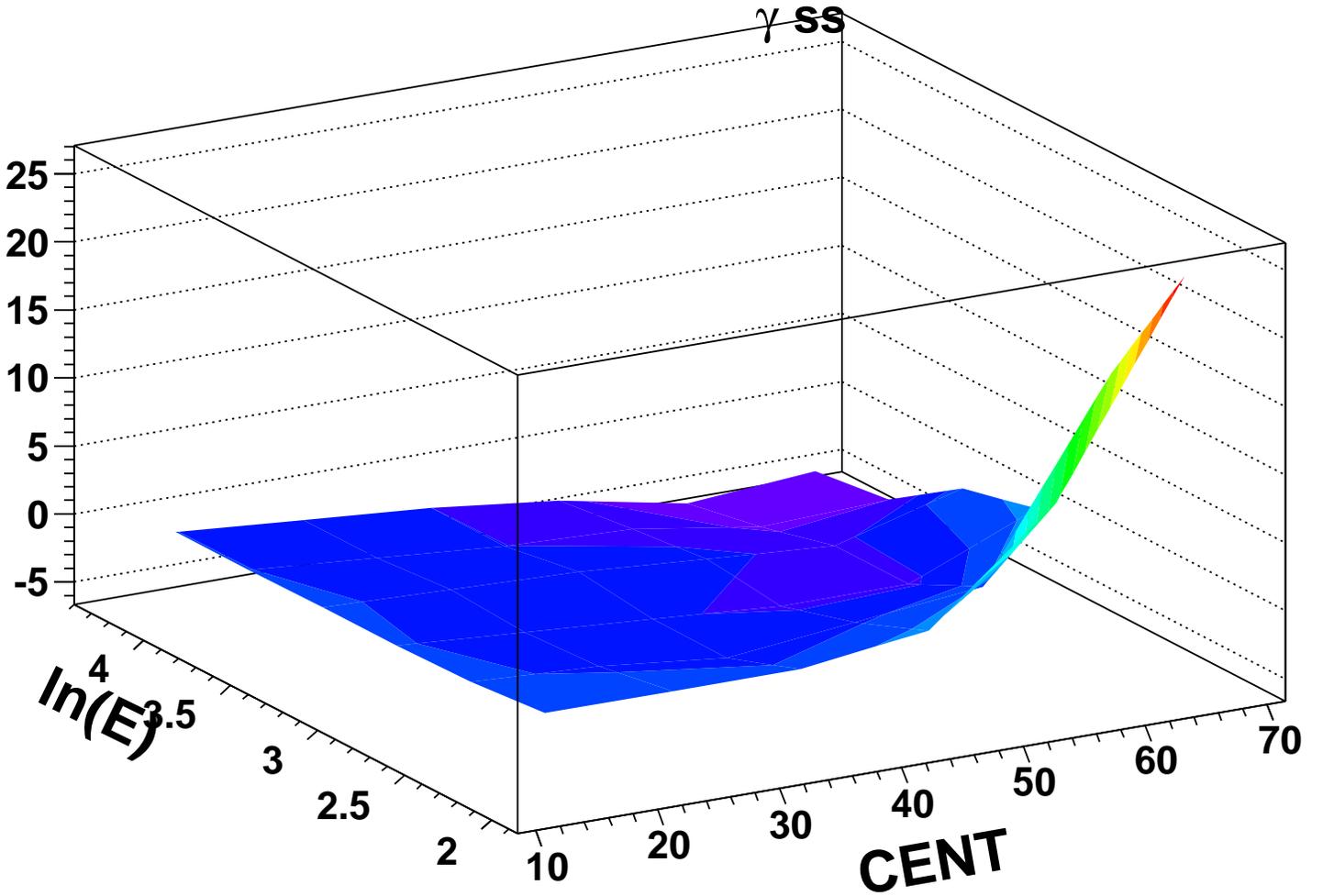}}
\end{center}
\vspace{2pt}
\caption{ (Color) Scatter plot of the correlation(units of $10^{-4}$) of same 
sign charge pairs $\langle cos(\phi_1 + \phi_2 - 2\Psi_{RP}) \rangle$ 
($\gamma$) calculated from Au-Au collisions with acceptance cuts of 0.15 $<$ 
$p_t$ $<$ 2 GeV/c and $|\eta|$ $<$ 1.0. The two axes are centrality vs beam 
energy. The beam energy is plotted as log($\sqrt{s_{NN}}$) with data from 
STAR\cite{STARBES} experiment at RHIC(see text). log($\sqrt{s_{NN}}$ = 
64.4 GeV) = 4.1 and log($\sqrt{s_{NN}}$ = 7.7 GeV) = 2.0. Centrality ranges 
from most central collisions at 10\%, while most peripheral at 70\%.}
\label{fig1}
\end{figure}

\begin{figure}
\begin{center}
\mbox{
   \epsfysize 8.0in
   \epsfbox{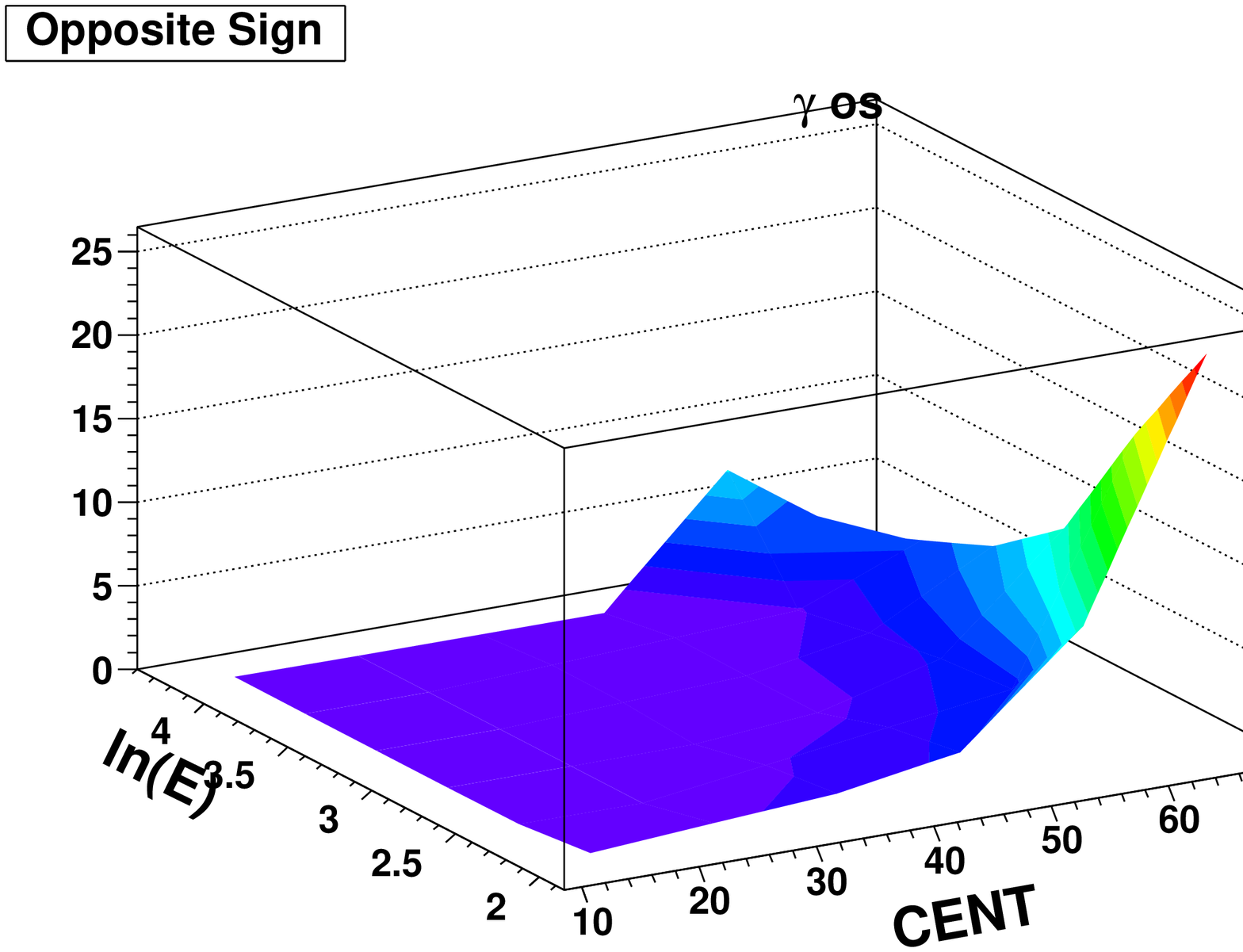}}
\end{center}
\vspace{2pt}
\caption{ (Color) Scatter plot of the correlation(units of $10^{-4}$)
of opposite sign charge pairs $\langle cos(\phi_1 + \phi_2 - 2\Psi_{RP}) 
\rangle$ ($\gamma$) calculated from Au-Au collisions with acceptance cuts of 
0.15 $<$ $p_t$ $<$ 2 GeV/c and $|\eta|$ $<$ 1.0. The two axes are centrality 
vs beam energy. The beam energy is plotted as log($\sqrt{s_{NN}}$) with data 
from STAR\cite{STARBES} experiment at RHIC(see text). log($\sqrt{s_{NN}}$ = 
64.4 GeV) = 4.1 and log($\sqrt{s_{NN}}$ = 7.7 GeV) = 2.0. Centrality ranges 
from most central collisions at 10\%, while most peripheral at 70\%.}
\label{fig2}
\end{figure}
 
\begin{figure}
\begin{center}
\mbox{
   \epsfysize 8.0in
   \epsfbox{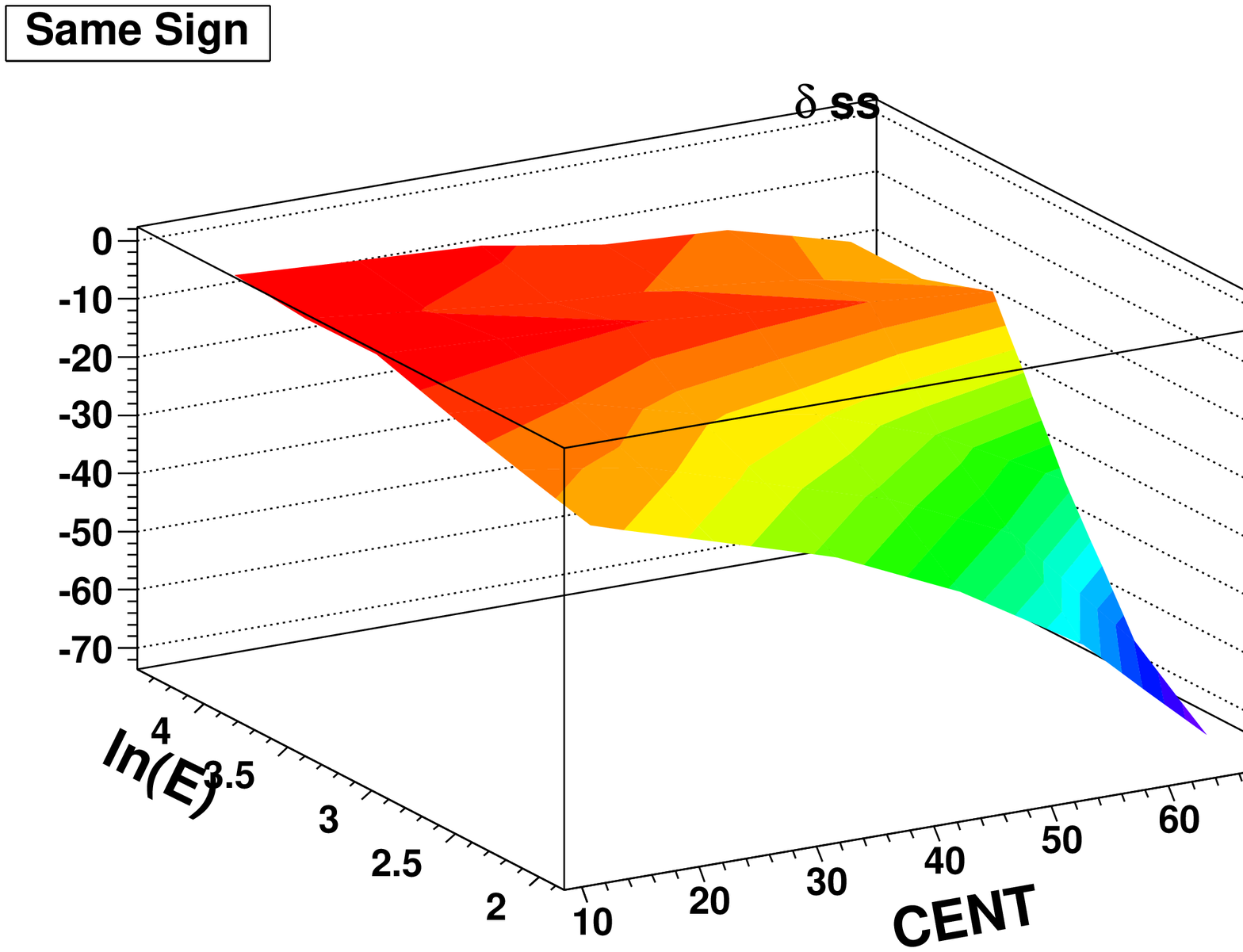}}
\end{center}
\vspace{2pt}
\caption{ (Color) Scatter plot of the correlation(units of $10^{-4}$) of same 
sign charge pairs $\langle cos(\phi_1 - \phi_2) \rangle$ ($\delta$) calculated 
from Au-Au collisions with acceptance cuts of 0.15 $<$ $p_t$ $<$ 2 GeV/c and 
$|\eta|$ $<$ 1.0. The two axes are centrality vs beam energy. The beam energy
is plotted as log($\sqrt{s_{NN}}$) with data from STAR\cite{STARBES} experiment
at RHIC(see text). log($\sqrt{s_{NN}}$ = 64.4 GeV) = 4.1 and 
log($\sqrt{s_{NN}}$ = 7.7 GeV) = 2.0. Centrality ranges from most 
central collisions at 10\%, while most peripheral at 70\%.}
\label{fig3}
\end{figure}
 
\begin{figure}
\begin{center}
\mbox{
   \epsfysize 8.0in
   \epsfbox{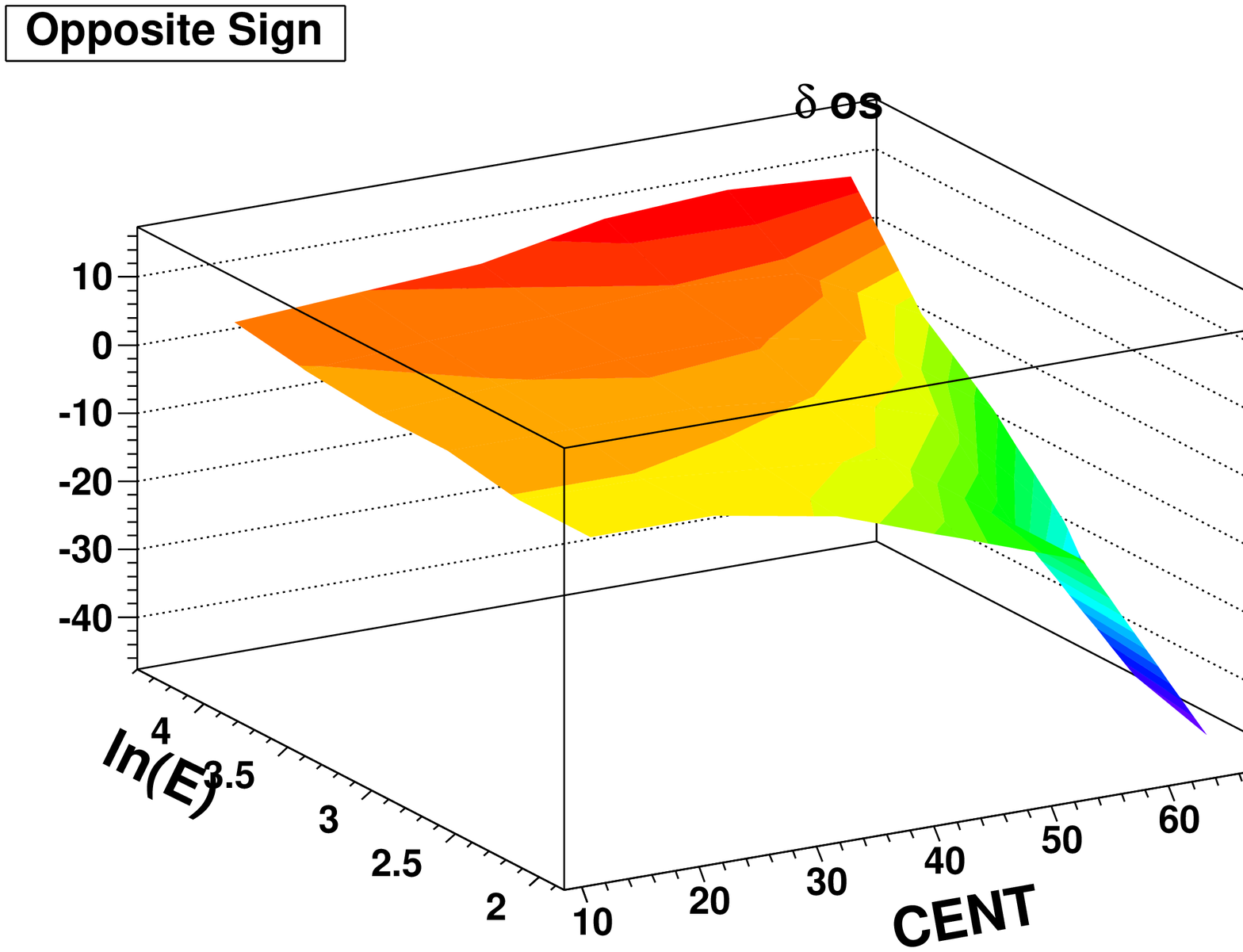}}
\end{center}
\vspace{2pt}
\caption{ (Color) Scatter plot of the correlation(units of $10^{-4}$) 
of opposite sign charge pairs $\langle cos(\phi_1 - \phi_2) \rangle$ 
($\delta$) calculated from Au-Au collisions with acceptance cuts of 0.15 $<$ 
$p_t$ $<$ 2 GeV/c and $|\eta|$ $<$ 1.0. The two axes are centrality vs beam 
energy. The beam energy is plotted as log($\sqrt{s_{NN}}$) with data from 
STAR\cite{STARBES} experiment at RHIC(see text). log($\sqrt{s_{NN}}$ = 
64.4 GeV) = 4.1 and log($\sqrt{s_{NN}}$ = 7.7 GeV) = 2.0. Centrality ranges 
from most central collisions at 10\%, while most peripheral at 70\%.}
\label{fig4}
\end{figure}
 
\begin{figure}
\begin{center}
\mbox{
   \epsfysize 8.0in
   \epsfbox{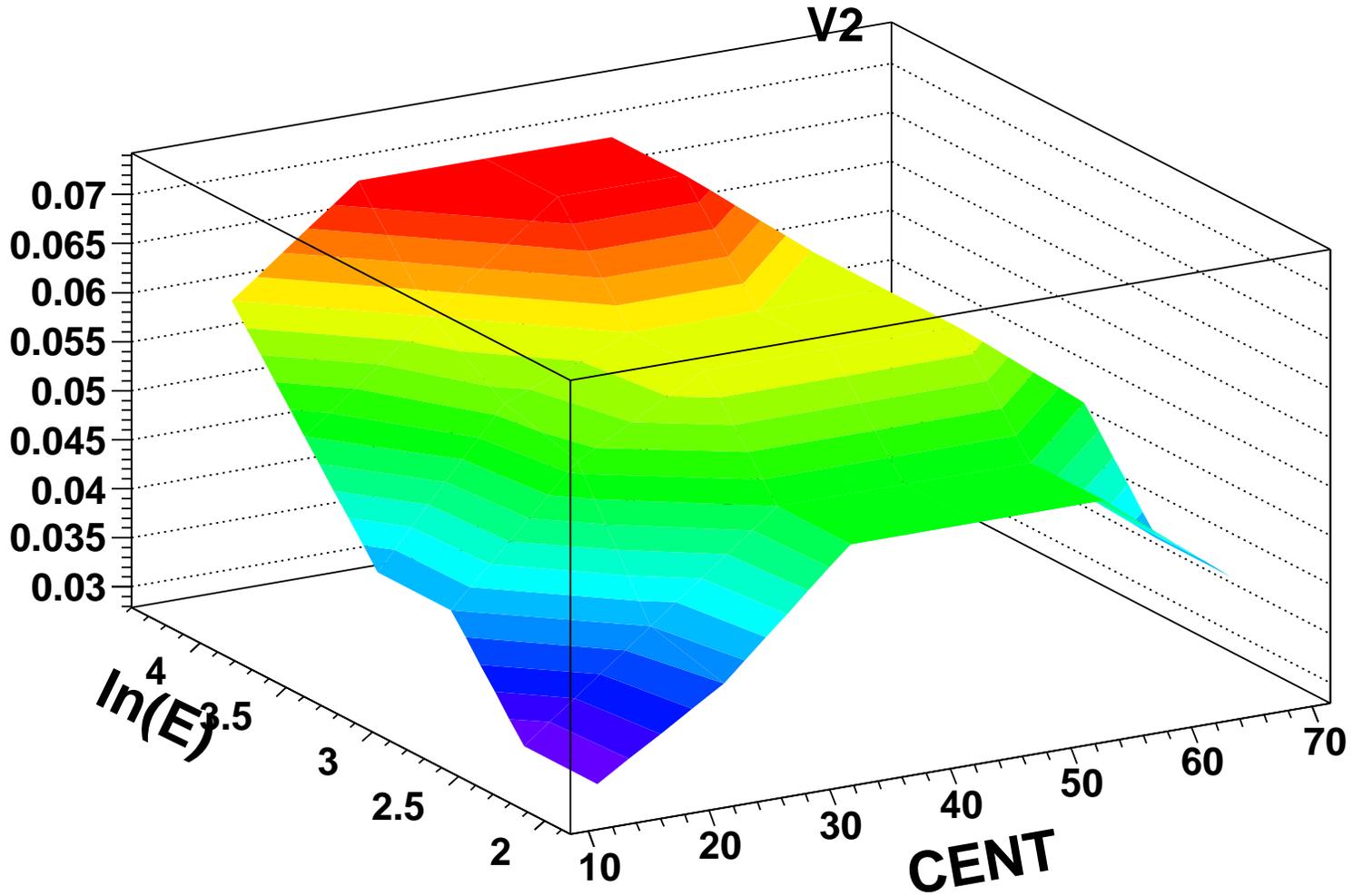}}
\end{center}
\vspace{2pt}
\caption{ (Color) Scatter plot of the average $v_2$ calculated from 
Au-Au collisions with acceptance cuts of 0.15 $<$ $p_t$ $<$ 2 GeV/c and 
$|\eta|$ $<$ 1.0. The two axes are centrality vs beam energy. The beam energy
is plotted as log($\sqrt{s_{NN}}$) with data from STAR\cite{STARBES} experiment
at RHIC(see text). log($\sqrt{s_{NN}}$ = 64.4 GeV) = 4.1 and 
log($\sqrt{s_{NN}}$ = 7.7 GeV) = 2.0. Centrality ranges from most 
central collisions at 10\%, while most peripheral at 70\%.}
\label{fig5}
\end{figure}

\section{Reaction Plane Dependent and Reaction Plane Independent Correlations}

Using the $\gamma$ and $\delta$ measurements we can define reaction plane 
dependent(H) and reaction plane independent(F) correlations. The 
definitions\cite{STARBES} from Ref.\cite{notes} are written as
\begin{equation}
\gamma = v_2 F - H,
\end{equation}
and
\begin{equation}
\delta = F + H.
\end{equation}

We can solve for H by using equations 4 and 5 obtaining
\begin{equation}
H = (v_2 \delta - \gamma)/( 1 + v_2).
\end{equation}
Thus F is given by
\begin{equation}
F = \delta - H.
\end{equation}

\subsection{Charge Dependent and Charge Independent Amplitudes}

The CME is a charge dependent amplitude. The CME amplitude has an opposite 
sign depending on whether we consider same sign charge pairs or opposite sign 
charge pairs. Also the CME is a reaction plane dependent amplitude acting
on pairs which are moving along the B field perpendicular to the reaction 
plane. For this analysis we will assume that the CME is the only such effect
at work when we isolate the reaction plane dependent(H) and charge dependent 
amplitude.  
\begin{equation}
\rm HCME = \it (H_{ss} - H_{os}) \rm/2,
\end{equation}
where $H_{ss}$ is the same sign charge pairs reaction plane dependent 
correlation and $H_{os}$ is the opposite sign charge pairs reaction plane 
dependent correlation(see Figure 6). At low log($\sqrt{s_{NN}}$ = 7.7 GeV) = 
2.0 HCME is very small, while at log($\sqrt{s_{NN}}$ = 64.4 GeV) = 4.1 HCME
becomes large $\sim$.0003. The highest value of HCME $\sim$.0005 is an 
isolated point at top energies(see rotated plot Figure 7). At all energies 
the HCME is small for central collisions(note B field is small for central 
collisions).

We can define three other amplitudes HCI(reaction plane dependent amplitude
with no charge dependence), FCD(reaction plane independent amplitude
with a charge dependence) and FCI(reaction plane independent amplitude
with no charge dependence). 
\begin{equation}
\rm HCI = \it (H_{ss} + H_{os}) \rm/2,
\end{equation}
\begin{equation}
\rm FCD = \it (F_{ss} - F_{os}) \rm/2,
\end{equation}
and
\begin{equation}
\rm FCI = \it (F_{ss} + F_{os}) \rm/2.
\end{equation}

\begin{figure}
\begin{center}
\mbox{
   \epsfysize 7.0in
   \epsfbox{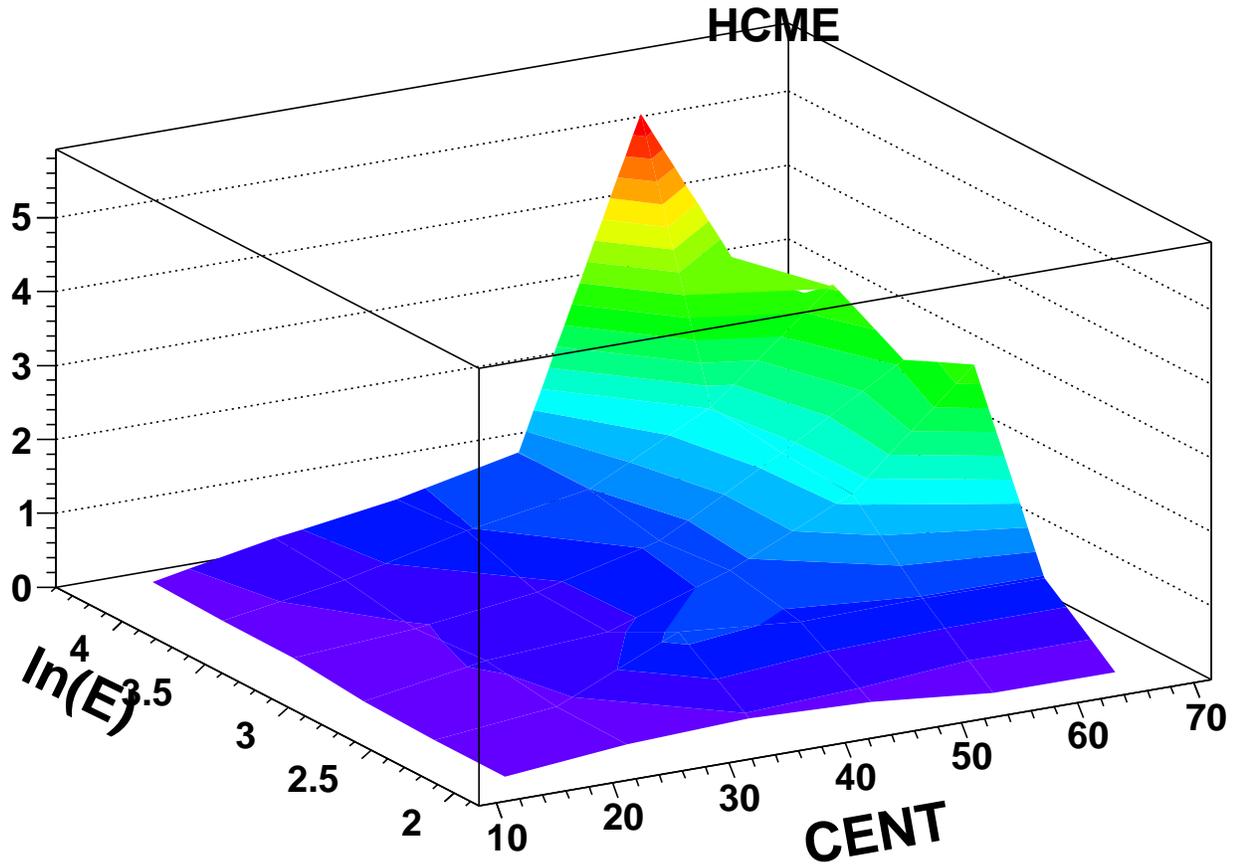}}
\end{center}
\vspace{2pt}
\caption{ The HCME is assumed to dominate this charge sign dependent reaction 
plane dependent amplitude(units of $10^{-4}$). The amplitude is calculated from
Au-Au collisions with acceptance cuts of 0.15 $<$ $p_t$ $<$ 2 GeV/c and 
$|\eta|$ $<$ 1.0. The two axes are centrality vs beam energy. The beam energy 
is plotted as log($\sqrt{s_{NN}}$) with data from STAR\cite{STARBES} experiment
 at RHIC(see text). log($\sqrt{s_{NN}}$) = 64.4 GeV) = 4.1 and 
log($\sqrt{s_{NN}}$ = 7.7 GeV) = 2.0. Centrality ranges from most central 
collisions at 10\%, while most peripheral at 70\%. HCME becomes large 
$\sim$.0003. The highest value of HCME $\sim$.0005 is an isolated at top 
energies(see rotated plot Figure 7). At all energies the HCME is small for 
central collisions(note B field is small for central collisions).}
\label{fig6}
\end{figure}

\begin{figure}
\begin{center}
\mbox{
   \epsfysize 8.0in
   \epsfbox{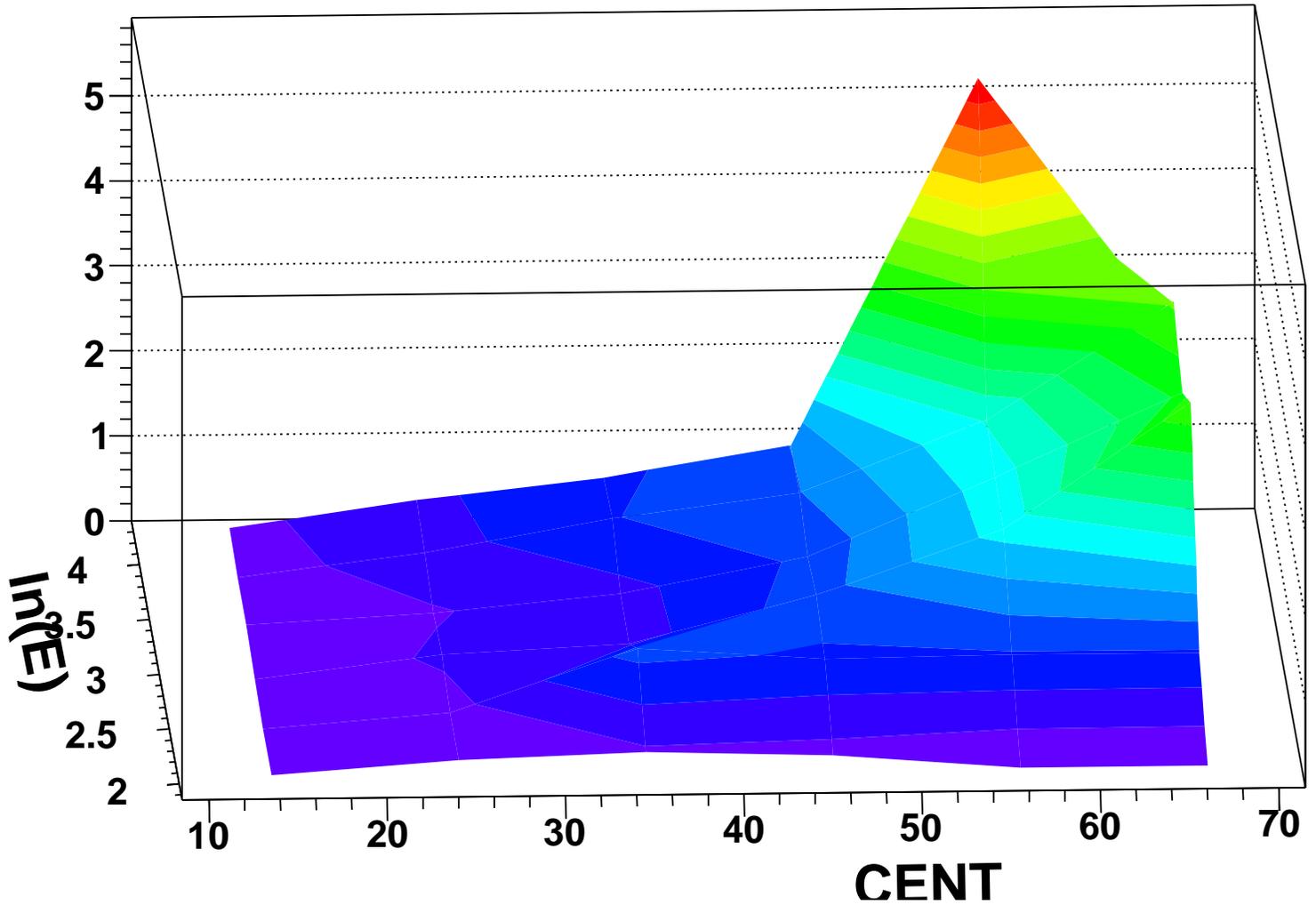}}
\end{center}
\vspace{2pt}
\caption{ The HCME has been rotated so one can see that HCME $\sim$.0005 is an 
isolated value at top energies while sitting on a plateau of $\sim$.0003 in
value. The amplitude(units of $10^{-4}$) is calculated from Au-Au collisions 
with acceptance cuts of 0.15 $<$ $p_t$ $<$ 2 GeV/c and $|\eta|$ $<$ 1.0. The 
two axes are centrality vs beam energy. The beam energy is plotted as 
log($\sqrt{s_{NN}}$) with data from STAR\cite{STARBES} experiment at 
RHIC(see text). log($\sqrt{s_{NN}}$ = 64.4 GeV) = 4.1 and log($\sqrt{s_{NN}}$ 
= 7.7 GeV) = 2.0. Centrality ranges from most central collisions at 10\%, 
while most peripheral at 70\%}. 
\label{fig7}
\end{figure}

The HCI is dominate at low beam energy and most peripheral. HCI is a charge 
sign independent reaction plane dependent amplitude(units of $10^{-4}$) as
defined above. At lower beam energies and most peripheral the shower of
charge particles from the Au-Au collision defines the reaction plane.
This amplitude is negative which is driven by the back to back nature of the
shower of particles due to momentum conservation. At the lowest beam energy 
the amplitude scales as 1/multiplicity(see Figure 8).

The FCD amplitude is a very complex two particle correlation where the 
same charge sign pair correlation has the opposite charge sign pair correlation
subtracted from it. At the highest beam energies opposite sign particles are 
correlated with each other moving together and creating a positive 
sign(see Figure 4 and Ref.\cite{SS-OS}). This positive correlation is stronger 
in the most peripheral collisions. The same charge sign correlation is negative
arising from back to back correlation between like sign particles. The 
difference thus is an overall negative correlation which is strongest for the 
most peripheral collisions. When we plot the FCD in our usual scatter 
plot(see Figure 9) this high beam energy behavior is hidden by the more 
complicated action at lower beam energies. In Figure 10 we have rotated the 
plot so we can see this negative value at the most peripheral which decreases 
in absolute value as 1/multiplicity with centrality.

The FCI is dominate at low beam energy and most peripheral. FCI is a charge 
sign independent reaction plane independent amplitude(units of $10^{-4}$) as
defined above. At lower beam energies and most peripheral the shower of
charge particles from the Au-Au collision have a back to back nature
due to momentum conservation. The same charge sign pairs follow this basic
behavior. Opposite charge sign pairs go against this behavior and cancel out
this display of momentum conservation except for the most peripheral and lowest
beam energy where momentum conservation is the strongest effect giving the 
largest negative amplitude of $\sim$-.0035 in this analysis(see Figure 11). 

\begin{figure}
\begin{center}
\mbox{
   \epsfysize 7.0in
   \epsfbox{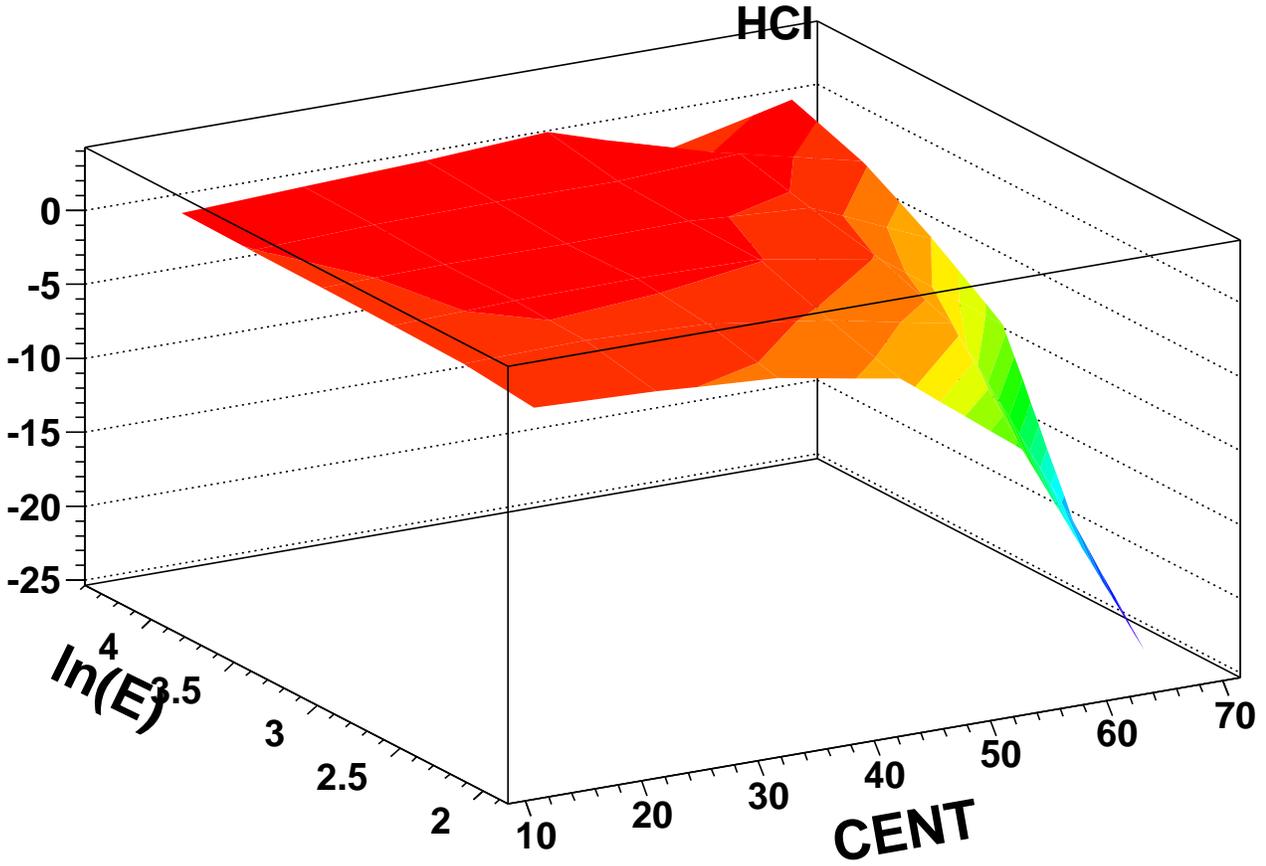}}
\end{center}
\vspace{2pt}
\caption{The HCI is dominate at low beam energy and most peripheral. HCI is a
charge sign independent reaction plane dependent amplitude(units of $10^{-4}$).
The amplitude is calculated from Au-Au collisions with acceptance cuts of 
0.15 $<$ $p_t$ $<$ 2 GeV/c and $|\eta|$ $<$ 1.0. The two axes are centrality 
vs beam energy. The beam energy is plotted as log($\sqrt{s_{NN}}$) with data 
from STAR\cite{STARBES} experiment at RHIC(see text). log($\sqrt{s_{NN}}$ = 
64.4 GeV) = 4.1 and log($\sqrt{s_{NN}}$ = 7.7 GeV) = 2.0. Centrality ranges 
from most central collisions at 10\%, while most peripheral at 70\%. HCI has 
a large negative value $\sim$-.0025. This is a back to back pair correlation 
of momentum conservation. At the lowest beam energy the scaling is one of 
1/multiplicity.}
\label{fig8}
\end{figure}

\begin{figure}
\begin{center}
\mbox{
   \epsfysize 7.0in
   \epsfbox{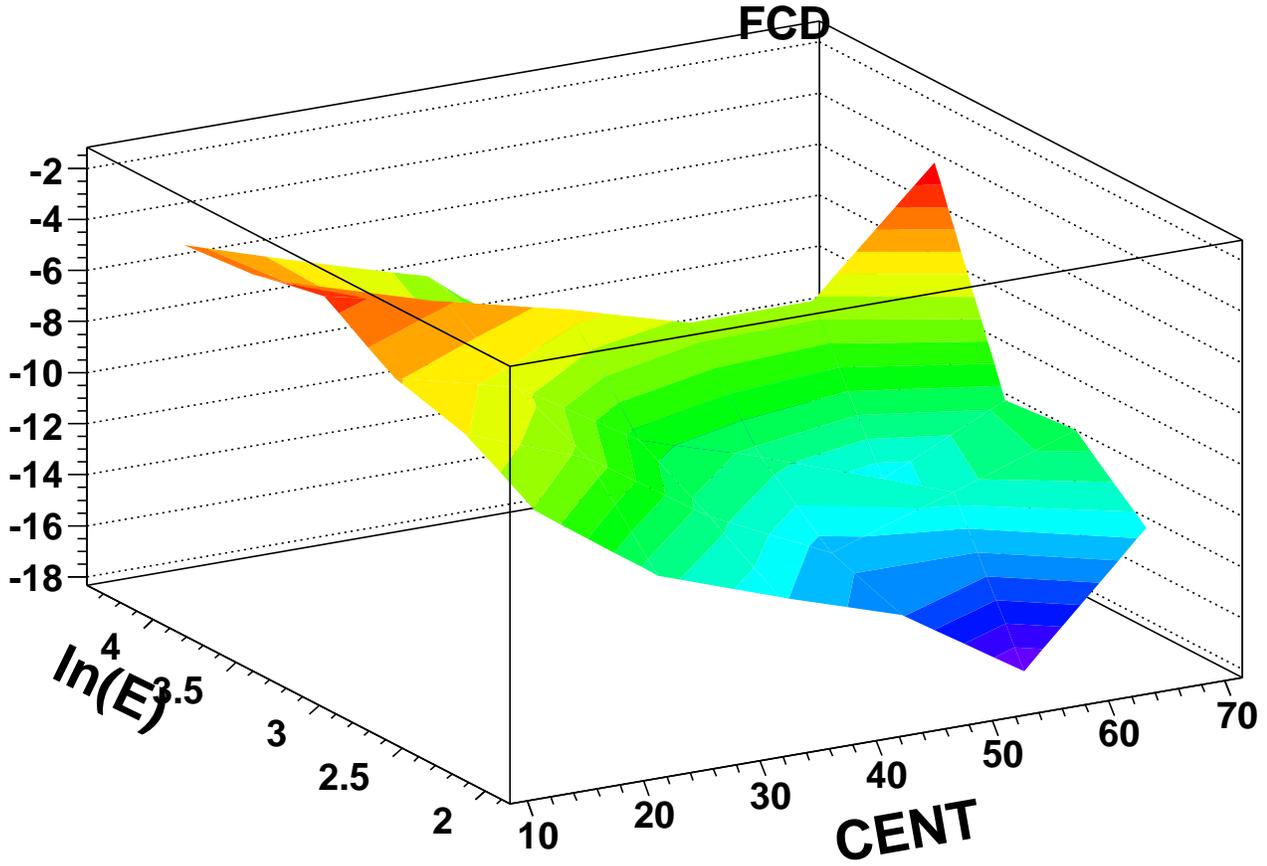}}
\end{center}
\vspace{2pt}
\caption{ The FCD amplitude is a very complex two particle pair correlation 
where the same charge sign pair correlation has the opposite charge sign pair 
correlation subtracted from it. When we plot the FCD in our usual scatter plot 
the high beam energy behavior is hidden by the more complicated action at lower
beam energies. In Figure 10 we have rotated the plot to so one can see the 
hidden behavior. FCD is a charge sign dependent reaction plane independent 
amplitude(units of $10^{-4}$). The amplitude is calculated from Au-Au 
collisions with acceptance cuts of 0.15 $<$ $p_t$ $<$ 2 GeV/c and $|\eta|$ 
$<$ 1.0. The two axes are centrality vs beam energy. The beam energy is 
plotted as log($\sqrt{s_{NN}}$) with data from STAR\cite{STARBES} experiment 
at RHIC(see text). log($\sqrt{s_{NN}}$ = 64.4 GeV) = 4.1 and 
log($\sqrt{s_{NN}}$ = 7.7 GeV) = 2.0. Centrality ranges from most 
central collisions at 10\%, while most peripheral at 70\%. FCD has a large 
negative value $\sim$-.0018 for lowest beam energy and centrality of 55\%.}
\label{fig9}
\end{figure}

\begin{figure}
\begin{center}
\mbox{
   \epsfysize 6.0in
   \epsfbox{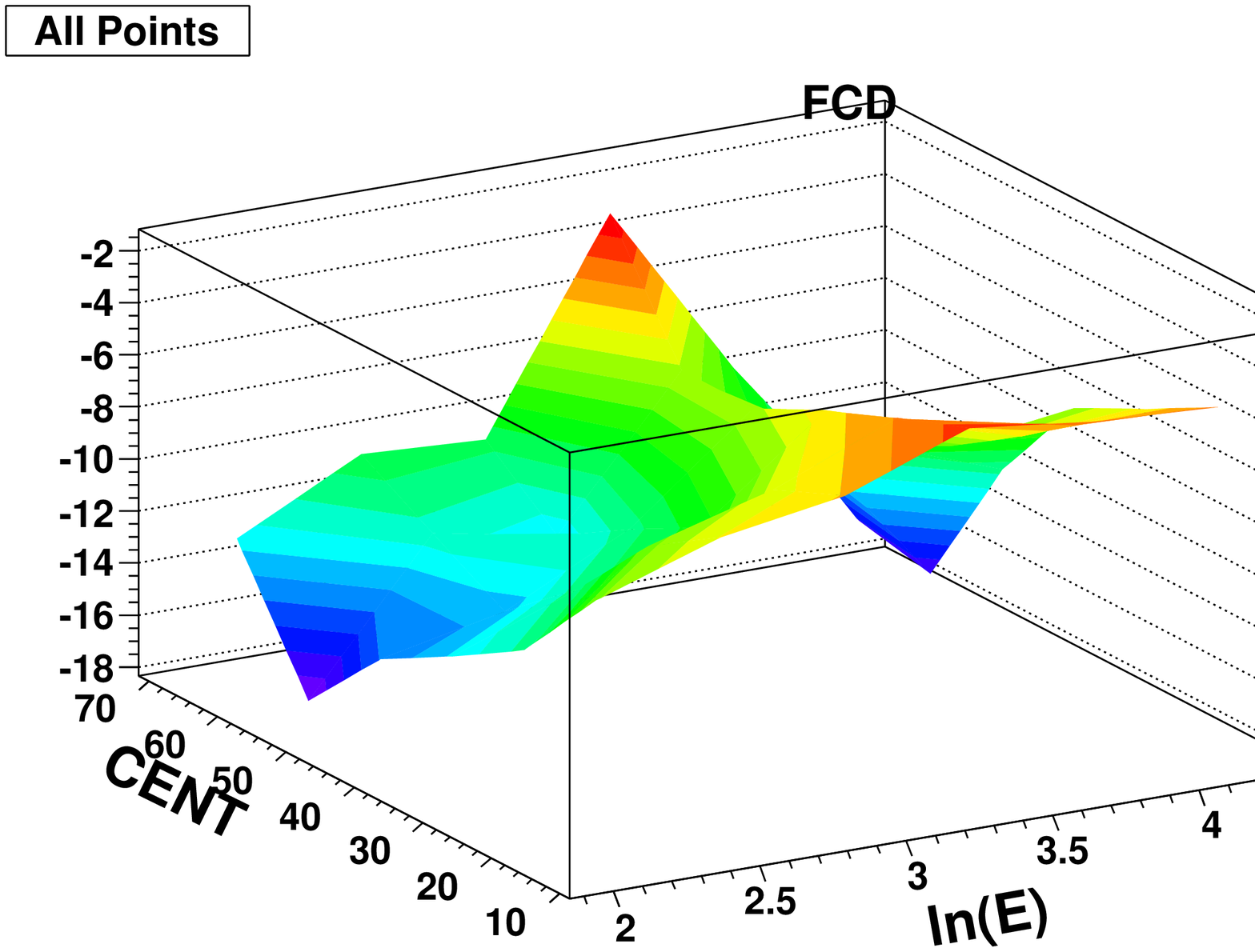}}
\end{center}
\vspace{2pt}
\caption{ As part of the FCD amplitude at the highest beam energies the 
opposite sign particles pairs are correlated with each other moving together 
and creating a positive sign(see Figure 4 and Ref.\cite{SS-OS}). This positive 
correlation is stronger in the most peripheral collisions but has a negative
contribution to FCD. The same charge sign pair correlation is negative arising
from back to back correlation between like sign particles and also makes the 
FCD more negative for peripheral collisions. Thus this overall negative 
correlation is strongest for the most peripheral collisions($\sim$-.0018). 
When we plot the FCD in our usual scatter plot(see Figure 9) this high beam 
energy behavior is hidden by the more complicated action at lower beam 
energies. In this figure we have rotated the plot so we can see this negative 
value at the most peripheral which decreases in absolute value as 
1/multiplicity with centrality. The FCD is a charge sign dependent reaction 
plane independent amplitude(units of $10^{-4}$). The amplitude is calculated 
from Au-Au collisions with acceptance cuts of 0.15 $<$ $p_t$ $<$ 2 GeV/c and 
$|\eta|$ $<$ 1.0. The two axes are centrality vs beam energy. The beam energy 
is plotted as log($\sqrt{s_{NN}}$) with data from STAR\cite{STARBES} experiment
at RHIC(see text). log($\sqrt{s_{NN}}$ = 64.4 GeV) = 4.1 and 
log($\sqrt{s_{NN}}$ = 7.7 GeV) = 2.0. Centrality ranges from most central 
collisions at 10\%, while most peripheral at 70\%.}
\label{fig10}
\end{figure}

\begin{figure}
\begin{center}
\mbox{
   \epsfysize 7.0in
   \epsfbox{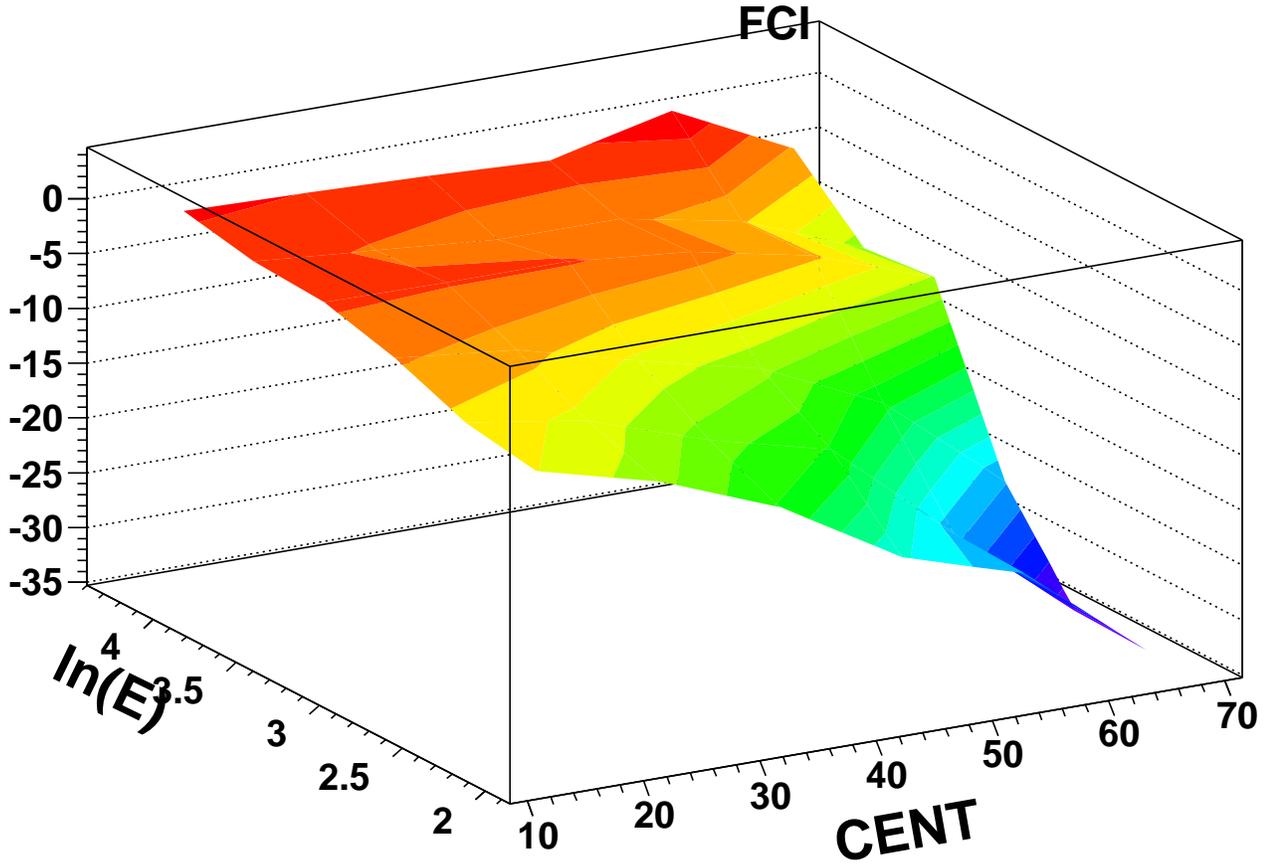}}
\end{center}
\vspace{2pt}
\caption{ The FCI is dominate at low beam energy and most peripheral. FCI is a 
charge sign independent reaction plane independent 
amplitude(units of $10^{-4}$). At lower beam energies and most peripheral the 
shower of charge particles from the Au-Au collision have a back to back nature 
due to momentum conservation. The same charge sign pairs follow this basic 
behavior. Opposite charge sign pairs go against this behavior and cancel out 
this display of momentum conservation except for the most peripheral and lowest
beam energy where momentum conservation is the strongest effect giving 
$\sim$-.0035 the largest negative amplitude in this analysis. The amplitude is 
calculated from Au-Au collisions with acceptance cuts of 0.15 $<$ $p_t$ $<$ 2 
GeV/c and $|\eta|$ $<$ 1.0. The two axes are centrality vs beam energy. The 
beam energy is plotted as log($\sqrt{s_{NN}}$) with data from 
STAR\cite{STARBES} experiment at RHIC(see text). log($\sqrt{s_{NN}}$ = 64.4 
GeV) = 4.1 and log($\sqrt{s_{NN}}$ = 7.7 GeV) = 2.0. Centrality ranges from 
most central collisions at 10\%, while most peripheral at 70\%.}
\label{fig11}
\end{figure}

\begin{figure}
\begin{center}
\mbox{
   \epsfysize 6.0in
   \epsfbox{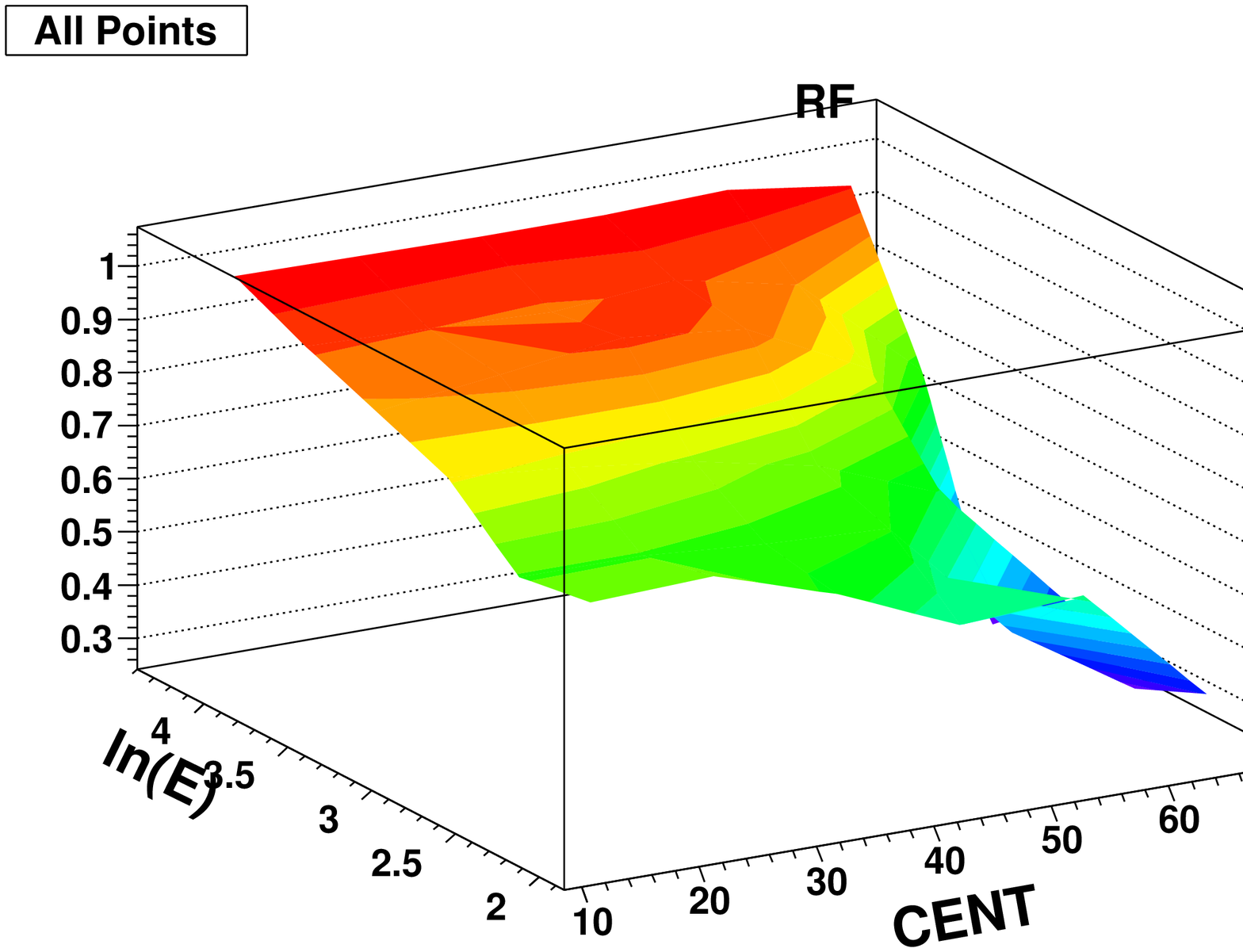}}
\end{center}
\vspace{2pt}
\caption{ The ratio(RF) of the charge dependent to the square root of the 
sum of the squares of the charge dependent(FCD) and the charge independent 
amplitudes(FCI). This ratio RF for reaction plane independent amplitudes 
for the higher beam energies the difference between the same charge sign 
pair correlation and the opposite charge sign pair correlation becomes vary
large leading to dominance of charge dependent amplitudes. The ratio is 
calculated from Au-Au collisions with acceptance cuts of 0.15 $<$ $p_t$ $<$ 
2 GeV/c and $|\eta|$ $<$ 1.0. The two axes are centrality vs beam energy. 
The beam energy is plotted as log($\sqrt{s_{NN}}$) with data from 
STAR\cite{STARBES} experiment at RHIC(see text). log($\sqrt{s_{NN}}$ = 
64.4 GeV) = 4.1 and log($\sqrt{s_{NN}}$ = 7.7 GeV) = 2.0. Centrality ranges 
from most central collisions at 10\%, while most peripheral at 70\%.}
\label{fig12}
\end{figure}

\section{Ratio Charge Dependent Amplitudes to Total}

For the higher beam energies the difference between the same charge sign 
pair correlation and the opposite charge sign pair correlation becomes vary
large leading to dominance of charge dependent amplitudes. We can express this
dominance by a ratio of the charge dependent to the square root of the sum
of the squares of the charge dependent and the charge independent amplitudes.
This ratio for the F or reaction plane independent amplitudes is given by
\begin{equation}
RF = FCD/\sqrt{FCD^2 + FCI^2}.
\end{equation}
RF is shown in Figure 12 where for high beam energy this ratio is 1.0 dropping
to 0.3 at the lowest energy and most peripheral.

The Chiral Magnetic Effect (HCME) is the dominate reaction plane and charge
dependent amplitude for peripheral collision where there will be a magnetic
field and high enough energy to have deconfined quarks. This ratio(RH) of the 
charge dependent to the square root of the sum of the squares of the charge 
dependent and the charge independent amplitudes is
\begin{equation}
RH = HCME/\sqrt{HCME^2 + HCI^2}.
\end{equation}
RH is shown in Figure 13 where the magnetic field is large and the energy is 
high enough for deconfined quarks the ratio is 1.0 dropping to zero at central 
events and low energies.

\section{Quenching of the CME}

Ref.\cite{warringa} has pointed out that P-odd domains on the surface of the
fireball omit same charge sign particles in the direction of the magnetic 
field. The particles that escape the surface would be of the same sign while 
the charge particles moving in the opposite direction would be of opposite 
sign. These particles would run into the fireball and be thermalized and loss
their direction (quenched). This implies that $H_{ss}$ would be unaffected
by quenching, thus
\begin{equation}
H_{ss} =  HCME + HCI.
\end{equation}
If we would consider a quenching factor q such that when q equals 1 we have
maximum quenching, $H_{os}$ becomes
\begin{equation}
H_{os} = -(1-q)HCME + HCI.
\end{equation}

Even though the CME is mainly in the same charge sign pairs the definition of 
the charge sign dependent reaction plane dependent amplitude remains the same. 
When there is quenching the relationship changes and the HCME becomes
\begin{equation}
(H_{ss} - H_{os})/2 = \left(2-q\over2\right) HCME.
\end{equation}
This causes in the maximum quenching case the HCME to be twice the value of the
charge sign dependent reaction plane dependent amplitude(see Figure 14).
the charge sign independent reaction plane dependent amplitude now picks up
a component of the CME,
\begin{equation}
(H_{ss} + H_{os})/2 = \left(q\over2\right) HCME + HCI.
\end{equation}
However this has a very small effect on this amplitude(see Figure 15). Finally
The ratio(RH) of the charge dependent to the square root of the sum of the 
squares of the charge dependent and the charge independent amplitudes is shown
in Figure 16. We see that this ratio appears to have a nice gaussian shape
which is consistent with there being quenching present in the reacting systems.
 
\begin{figure}
\begin{center}
\mbox{
   \epsfysize 7.0in
   \epsfbox{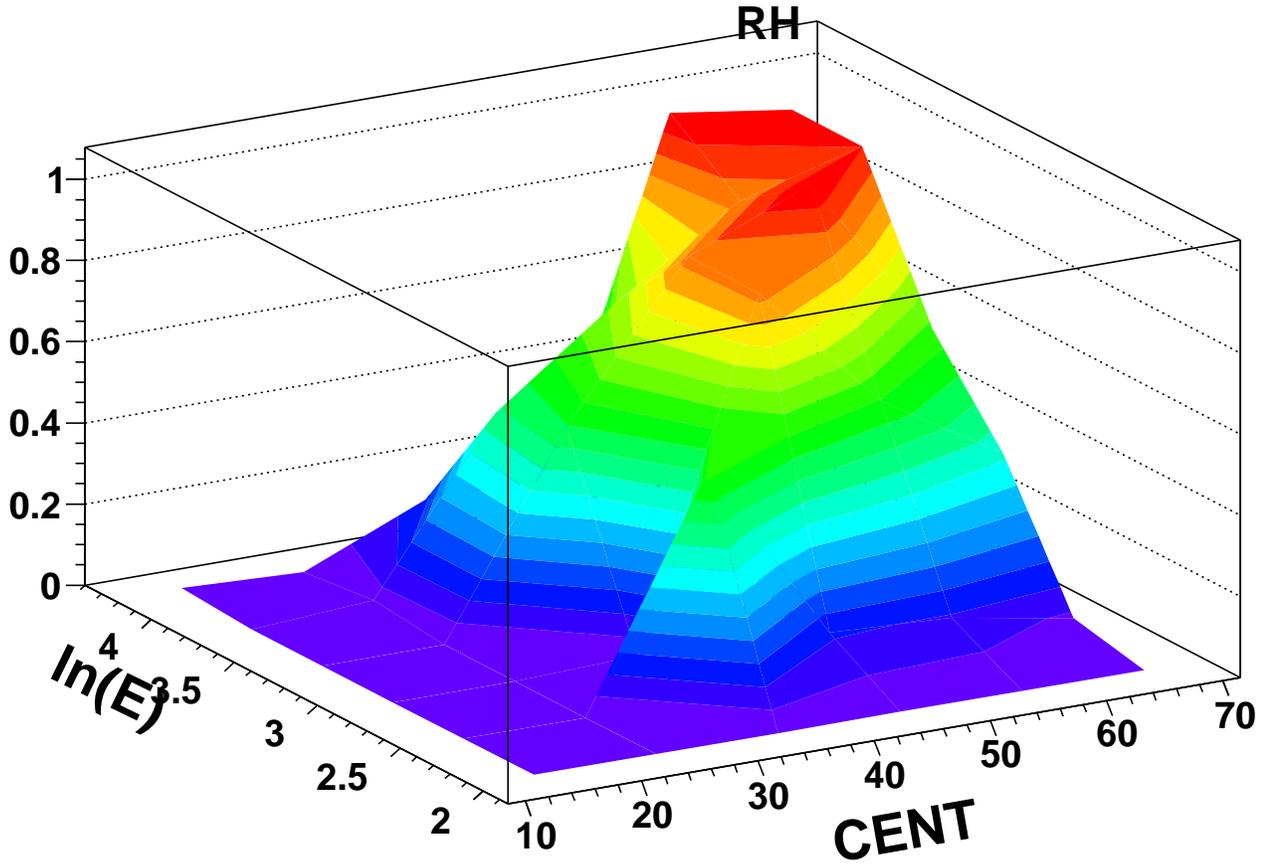}}
\end{center}
\vspace{2pt}
\caption{ The ratio(RH) of the charge and reaction plane dependent to the 
square root of the sum of the squares of the charge dependent(HCME) and 
the charge independent amplitudes(HCI). This ratio RH for reaction plane 
dependent amplitudes for the higher beam energies(deconfined quarks) and
peripheral collision (magnetic field) becomes vary large leading to dominance 
of charge dependent amplitudes. The ratio is calculated from Au-Au 
collisions with acceptance cuts of 0.15 $<$ $p_t$ $<$ 2 GeV/c and $|\eta|$ 
$<$ 1.0. The two axes are centrality vs beam energy. The beam energy is 
plotted as log($\sqrt{s_{NN}}$) with data from STAR\cite{STARBES} experiment 
at RHIC(see text). log($\sqrt{s_{NN}}$ = 64.4 GeV) = 4.1 and 
log($\sqrt{s_{NN}}$ = 7.7 GeV) = 2.0. Centrality ranges from most central 
collisions at 10\%, while most peripheral at 70\%.}
\label{fig13}
\end{figure}

\section{Summary and Discussion}

We use the STAR\cite{STARBES} correlation data in an analysis that can
separate the correlations up into reaction plane dependent and independent
parts with a further separation into charge dependent and charge independent
amplitudes. This gives us a clean separation of the The Chiral Magnetic Effect
into an amplitude HCME. This amplitude is isolated to top energies and
peripheral collisions where the magnetic is field large and quarks are
not confined. We also show that the idea of quenching is supported by the
data best expressed by the ratio RH. This ratio(RH) of the charge dependent to 
the square root of the sum of the squares of the charge dependent and the 
charge independent amplitudes which we show in Figure 16 appears to have a 
nice gaussian shape. This is consistent with there being quenching present in 
the reacting systems.

\begin{figure}
\begin{center}
\mbox{
   \epsfysize 7.0in
   \epsfbox{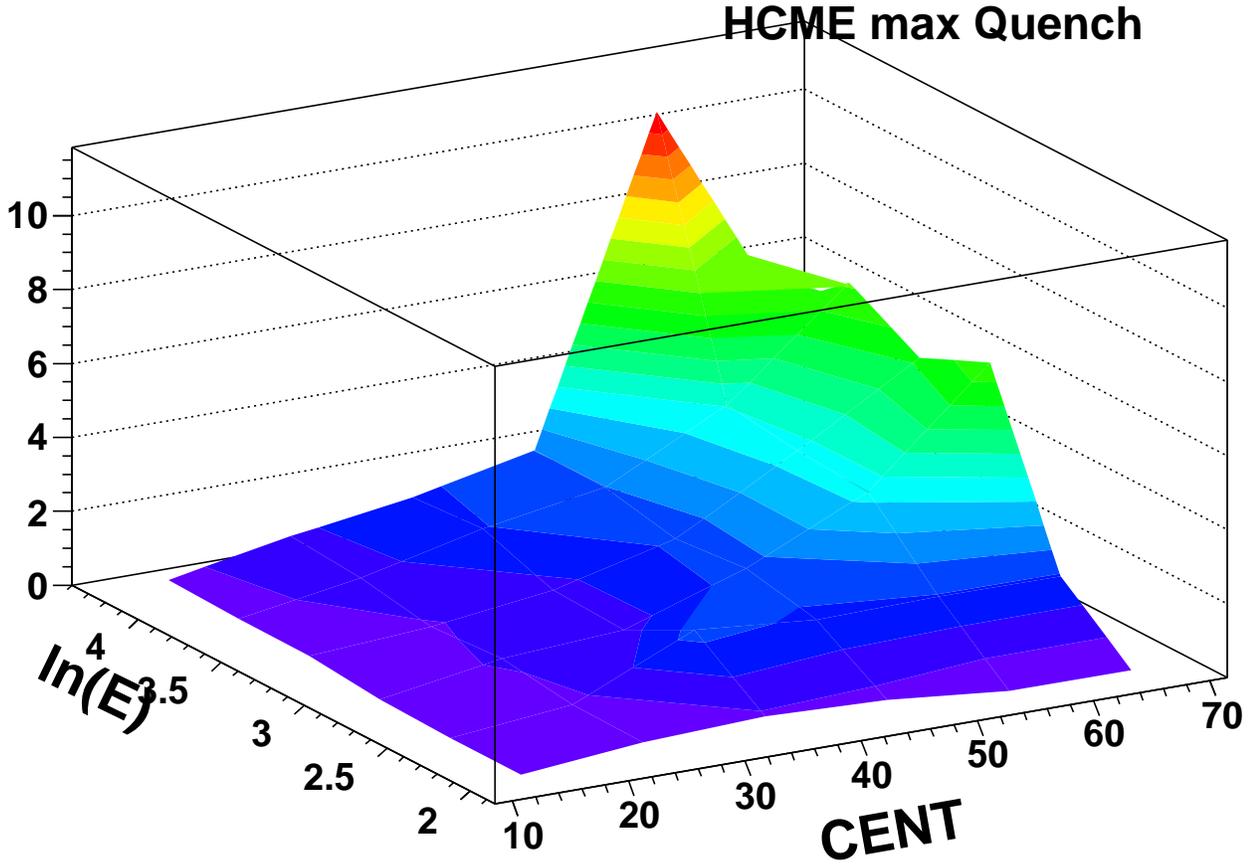}}
\end{center}
\vspace{2pt}
\caption{ The HCME is assumed to dominate this charge sign dependent reaction 
plane dependent amplitude(units of $10^{-4}$). Here we plot the HCME which 
would exist if there was maximum quenching. The amplitude is calculated from
Au-Au collisions with acceptance cuts of 0.15 $<$ $p_t$ $<$ 2 GeV/c and 
$|\eta|$ $<$ 1.0. The two axes are centrality vs beam energy. The beam energy 
is plotted as log($\sqrt{s_{NN}}$) with data from STAR\cite{STARBES} experiment
 at RHIC(see text). log($\sqrt{s_{NN}}$) = 64.4 GeV) = 4.1 and 
log($\sqrt{s_{NN}}$ = 7.7 GeV) = 2.0. Centrality ranges from most central 
collisions at 10\%, while most peripheral at 70\%. HCME becomes large 
$\sim$.0006. The highest value of HCME $\sim$.0010 is an isolated at top 
energies. At all energies central collisions the HCME is small(note B field is 
small at central collisions).}
\label{fig14}
\end{figure}
 
\begin{figure}
\begin{center}
\mbox{
   \epsfysize 7.0in
   \epsfbox{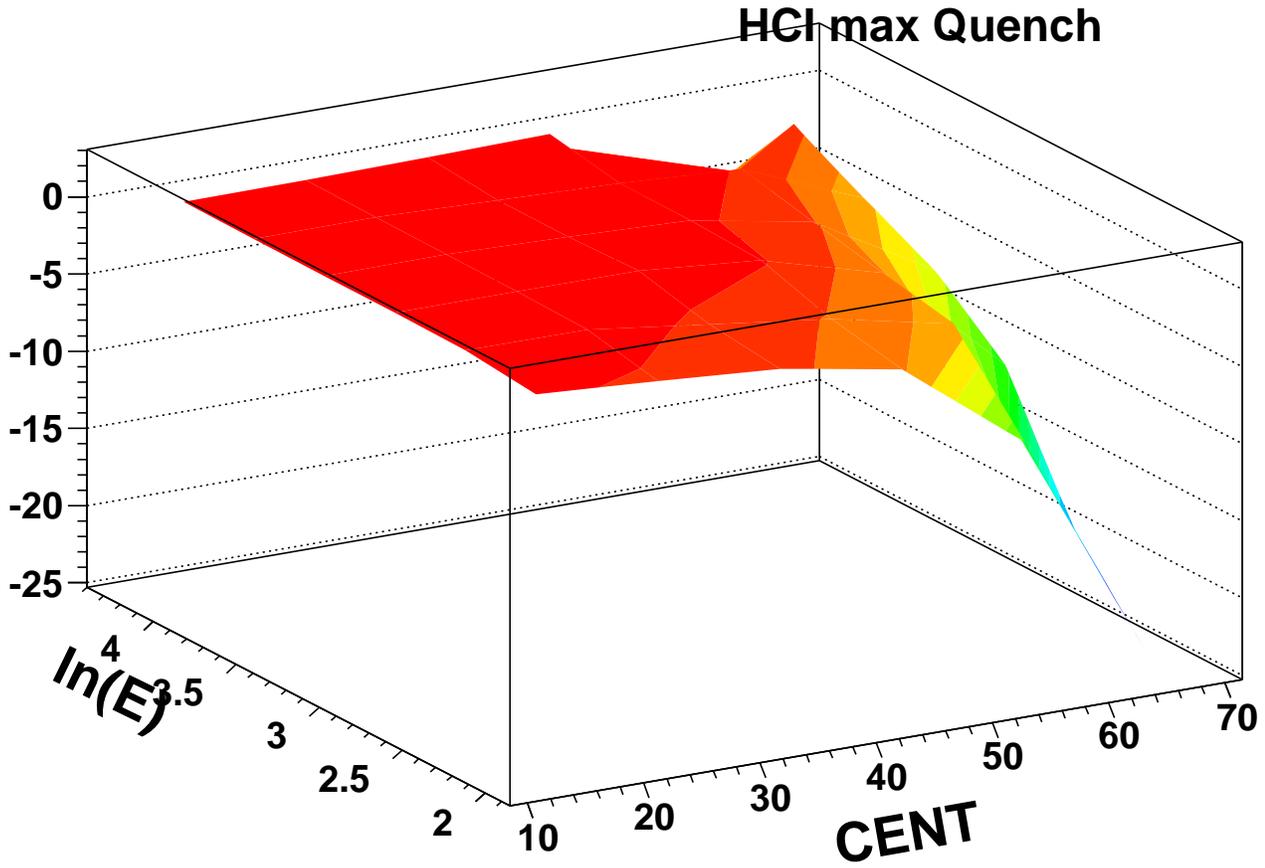}}
\end{center}
\vspace{2pt}
\caption{The HCI is dominate at low beam energy and most peripheral. Since this
is the region where the CME is small quenching has little effect on this 
amplitude. HCI is a charge sign independent reaction plane dependent 
amplitude(units of $10^{-4}$). The ratio is calculated from Au-Au 
collisions with acceptance cuts of 0.15 $<$ $p_t$ $<$ 2 GeV/c and $|\eta|$ 
$<$ 1.0. The two axes are centrality vs beam energy. The beam energy is 
plotted as log($\sqrt{s_{NN}}$) with data from STAR\cite{STARBES} experiment 
at RHIC(see text). log($\sqrt{s_{NN}}$ = 64.4 GeV) = 4.1 and 
log($\sqrt{s_{NN}}$ = 7.7 GeV) = 2.0. Centrality ranges from most central 
collisions at 10\%, while most peripheral at 70\%. HCI has a large negative 
value $\sim$-.0025. This is a back to back pair correlation of momentum 
conservation. At the lowest beam energy the scaling is one of 1/multiplicity.}
\label{fig15}
\end{figure}
\begin{figure}
\begin{center}
\mbox{
   \epsfysize 7.0in
   \epsfbox{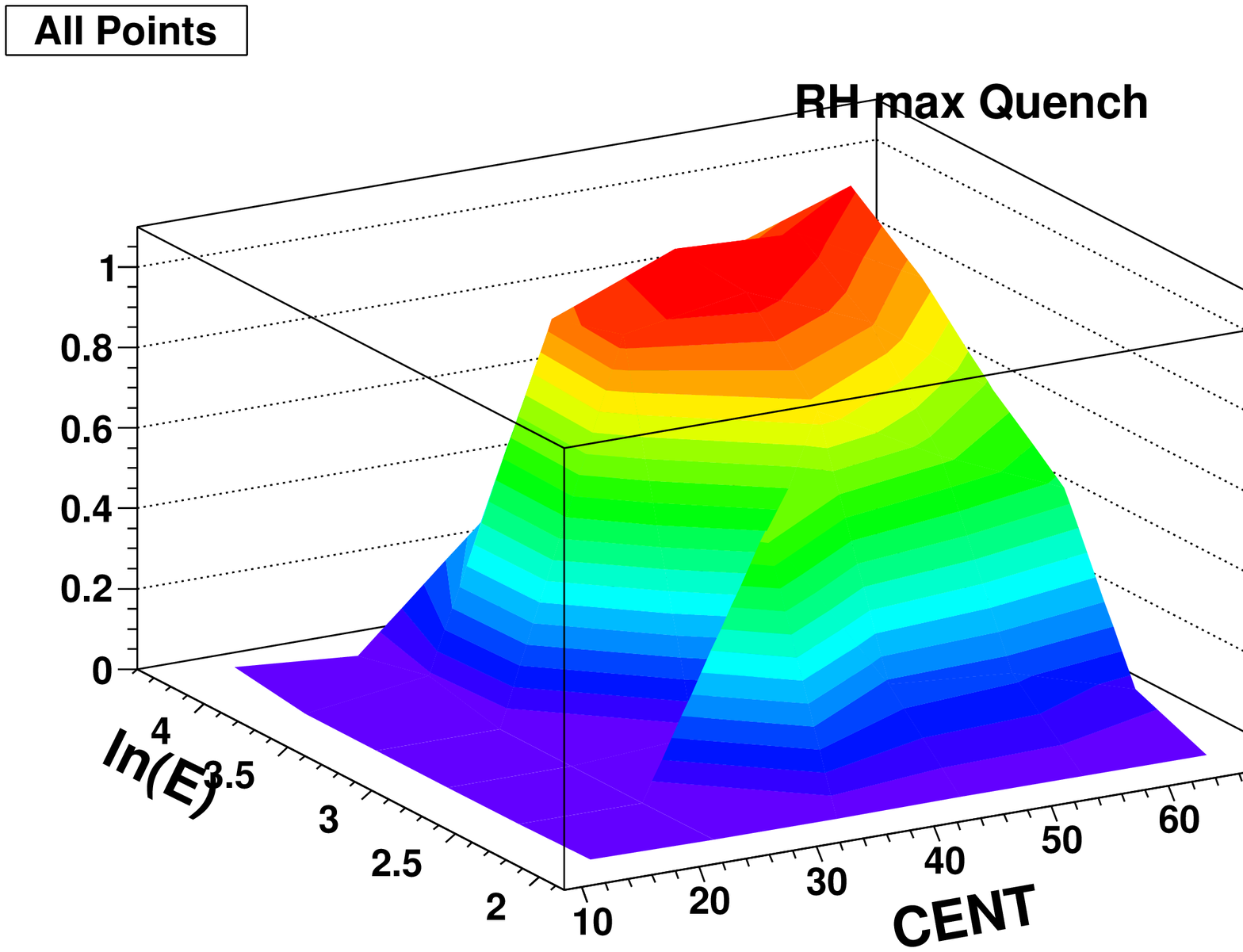}}
\end{center}
\vspace{2pt}
\caption{ The ratio(RH) of the charge and reaction plane dependent to the 
square root of the sum of the squares of the charge dependent(HCME) and 
the charge independent amplitudes(HCI). This ratio RH for reaction plane 
dependent amplitudes for the higher beam energies(deconfined quarks) and
peripheral collision (magnetic field) becomes vary large leading to dominance 
of charge dependent amplitudes. The amplitude is calculated from Au-Au 
collisions with acceptance cuts of 0.15 $<$ $p_t$ $<$ 2 GeV/c and $|\eta|$ 
$<$ 1.0. The two axes are centrality vs beam energy. The beam energy is 
plotted as log($\sqrt{s_{NN}}$) with data from STAR\cite{STARBES} experiment 
at RHIC(see text). log($\sqrt{s_{NN}}$ = 64.4 GeV) = 4.1 and 
log($\sqrt{s_{NN}}$ = 7.7 GeV) = 2.0. Centrality ranges from most central 
collisions at 10\%, while most peripheral at 70\%.}
\label{fig16}
\end{figure}

\section{Acknowledgments}

This research was supported by the U.S. Department of Energy under Contract No.
DE-AC02-98CH10886.

\end{document}